\documentclass{article}

\usepackage{arxiv}

\usepackage[utf8]{inputenc} 
\usepackage[T1]{fontenc}    
\usepackage{hyperref}       
\usepackage{url}            
\usepackage{booktabs}       
\usepackage{amsmath,amsfonts,amssymb}       
\usepackage{nicefrac}       
\usepackage{microtype}      
\usepackage{graphicx}
\usepackage[square,numbers]{natbib}
\usepackage{doi}

\title{Large Field-of-View Thermal Imaging via All-Silicon Meta-Optics}

\author{Anna Wirth-Singh$^{1,*}$\\
\And
Johannes E. Fr\"{o}ch$^{1,2}$\\
\And
Zheyi Han$^{2}$\\ 
\And 
Luocheng Huang$^{2}$\\
\And
Saswata Mukherjee$^{2}$\\
\And
Zhihao Zhou$^{2}$\\
\And
Zachary Coppens$^{3}$\\ 
\And
Karl F. B\"{o}hringer$^{2}$\\
\And
Arka Majumdar$^{1,2,**}$\\
\And
\\
$^{1}${Department of Physics, University of Washington, Seattle, WA 98195, USA}\\
$^{2}${Department of Electrical and Computer Engineering, University of Washington, Seattle, WA 98195, USA}\\
$^{3}${CFD Research Corporation, Huntsville, AL 35806, USA}\\
$^{*}$\textit{annaw77@uw.edu}\\
$^{**}$\textit{arka@uw.edu}
}

\hypersetup{
pdftitle={Large Field-of-View Thermal Imaging via All-Silicon Meta-Optics},
pdfsubject={optics},
pdfauthor={Anna Wirth-Singh},
pdfkeywords={meta-optics, field of view, thermal imaging},
}

\begin{document}
\maketitle

\begin{abstract}
A broad range of imaging and sensing technologies in the infrared require large Field-of-View (FoV) operation. To achieve this, traditional refractive systems often employ multiple elements to compensate for aberrations, which leads to excess size, weight, and cost. For many applications, including night vision eye-wear, air-borne surveillance, and autonomous navigation for unmanned aerial vehicles, size and weight are highly constrained. Sub-wavelength diffractive optics, also known as meta-optics, can dramatically reduce the size, weight, and cost of these imaging systems, as meta-optics are significantly thinner and lighter than traditional refractive lenses. Here, we demonstrate 80$^\circ$ FoV thermal imaging in the long-wavelength infrared regime (8-12 $\mu$m) using an all-silicon meta-optic with an entrance aperture and lens focal length of 1 cm. 
\end{abstract}

\keywords{Meta-optics, Field of View, Thermal Imaging}
\section{Introduction}

According to Planck's law of blackbody radiation, the emission from thermal sources with temperatures in the range of 0$^\circ$C to 100$^\circ$C has a maximum in its spectral intensity distribution in the long-wave infrared range (LWIR, 8-12 $\mu$m). Hence this spectral range is commonly used to achieve high temperature contrast with high detection efficiency for thermal imaging. Particularly for some night vision applications, this capability is critical to identify objects from their environment or to sense small temperature differences. In comparison, night vision capabilities that work in the near infrared range (800 nm - 2.5 $\mu$m) are typically not able to provide temperature contrast and rely on other illumination sources, such as the night sky or forward pointing IR LEDs/lasers, to work. This makes LWIR imaging attractive for a range of night vision applications, including many which are critical for defense and national security technologies \cite{Voll18,Estr03,Meem19,Kapl93}. 

At the same time, surveillance and imaging applications typically benefit from large Field-of-View (FoV), which allows a wide scene to be captured in a single frame. For comparison, human vision has a full field of view of approximately 120$^\circ$, and wide angle lenses are generally considered to be those with a field of view larger than 60$^\circ$ \cite{Laik80}. To achieve large FoV using traditional refractive optics,multiple lenses are often used to correct for Seidel aberrations, which leads to excess size and weight \cite{Kuml00,Bett05}. However, reduced size and weight are often crucial requirements for mobile or miniaturized systems, such as night vision goggles or airborne remote sensing. Weight reduction in particular is challenging for such systems because the optical elements necessary for the above mentioned applications typically require larger apertures to achieve low f-number and to compensate for low signal intensities to facilitate imaging over larger distances. 

Flat meta-optics have the potential to reduce the complex lens systems to fewer, thinner components, enabling more compact and lightweight optical solutions. Meta-optics are quasi periodic arrays of sub-wavelength scatterers which locally impart a phase shift on the incident light. By tuning the geometry and arrangement of the scatterers, the transmitted wavefront is locally modified to achieve a desired global operation, such as lensing, steering, encoding, or a combination of these \cite{Kama19}. The active thickness of meta-optics are on the order of a wavelength. Including the thickness of the substrate on which the structures are supported, a meta-optic is only 100 - 500 $\mu$m thick, which is orders of magnitude thinner than most thermal refractive optics.

The canonical hyperboloid lens phase profile can be used to obtain diffraction-limited resolution for normally incident light and small apertures. However, the resolution suffers severely from off-axis aberrations, thus limiting the FoV. For off-axis angles larger than  2$^\circ$, the point spread function (PSF) already becomes severely distorted \cite{Mart20}. By reducing the constraint of diffraction-limited quality, a quadratic phase profile can be used for large FoV imaging, and single-wavelength imaging at 532 nm using a 2 mm $(\sim 4000\lambda)$ aperture singlet lens of this phase profile has been demonstrated \cite{Mart20}. Despite the compactness and simplicity, a singlet lens suffers from stray background light because incident light from all angles interacts with all parts of the lens. 

Another approach towards realizing large FoV lenses is to use an external aperture to restrict incoming light such that beams of different angles of incidence interact with different parts of the meta-optic. Hence, each part of the meta-optic only needs to be optimized for a narrow range of angles of incidence. While reducing the diameter of the entrance aperture increases the spatial separation of the light, it also reduces the input light into the system and therefore the signal-to-noise ratio. This approach has been used to demonstrate very large FoV (near 170$^\circ$ at 5.2 $\mu$m wavelength), but for an extremely small entrance aperture of 1 mm $(\sim 200\lambda)$ \cite{Shal20}.

Following a similar approach, here we experimentally demonstrate an all-silicon large FoV meta-optic in the LWIR range. Our optical system has an aperture of 1 cm $(\sim 1000\lambda)$, whereas the meta-optic itself has an aperture of 3 cm. LWIR meta-optics, including all-silicon meta-optics, have previously been demonstrated \cite{Huan21,Sara23,Fanq18}, but wide FoV was not reported.multilevel diffractive all-silicon LWIR lenses with 46$^\circ$ FoV have been reported \cite{Kign21,Meem19}, but the fabrication of such multi-layer devices is more complex than that of single-layer meta-optics, and they suffer from higher-order diffraction.

\begin{figure}[h!]
\centering\includegraphics[width=12cm]{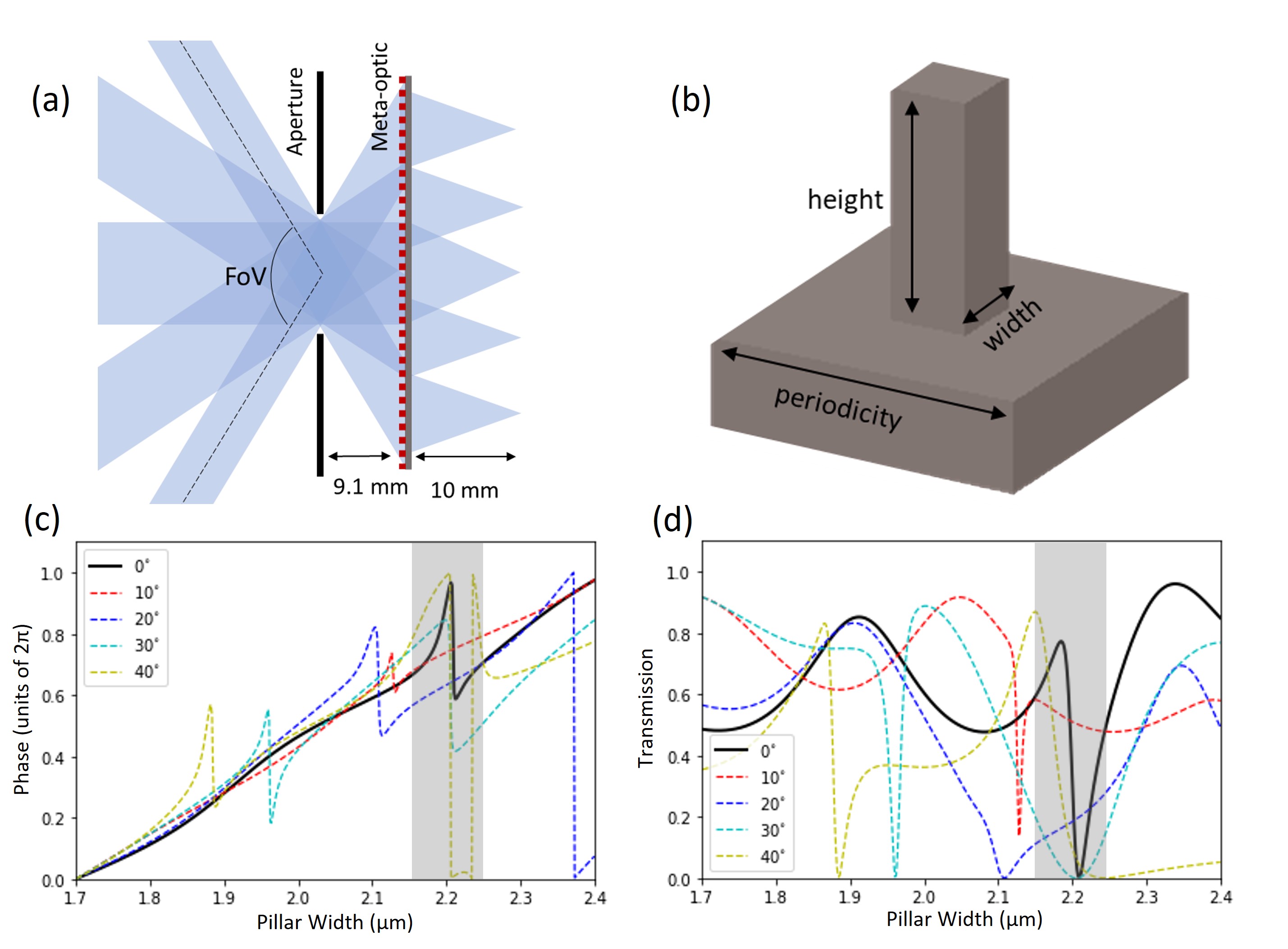}
\caption{Lens system design. (a) Schematic illustrating the entrance aperture and the meta-optic, with the path of light rays depicted by shaded blue regions. (b) Meta-atom schematic. The meta-optic consists of a periodic array (fixed periodicity) of square pillars of uniform height and variable width. (c) Scatterer phase and (d) transmission responses calculated using rigorous coupled wave analysis, for various angles of incidence. Pillar widths in the gray shaded region were excluded to ensure reasonable transmission.}
\label{Fig:Schematic}
\end{figure}

\section{Design}
The design of the lens system consists of determining the diameter and placement of the external aperture, the optimal phase mask of the meta-optic, and the geometry of the sub-wavelength scatterers to realize the phase mask. We designed the system for a fixed wavelength of 10 $\mu$m, a set entrance aperture diameter of 1 cm, and meta-optic focal length of 1 cm. First, to optimize the large-scale configuration of the aperture and the design of the phase mask, we used a commercial ray tracing software (Zemax OpticStudio) with an in-built optimization algorithm (root mean square) and merit function to minimize the spot size in the image plane for multiple angles of incidence. In detail, as shown in Figure \ref{Fig:Schematic}a, we assume an entrance aperture of 1 cm, chosen to balance the trade-off between light collection (requiring larger apertures) and achieving spatial separation of rays for different angles of incidence (favoring smaller apertures). The phase profile was assumed to be radially symmetric, of the form
\begin{equation}
\Phi = M\sum_{i=1}^{12} A_{i}\rho^{2i}    
\end{equation}
where $M$ is a normalization constant, $\rho$ is the normalized radial coordinate, and $A_{i}$ are polynomial coefficients. We defined input fields at incident angles of 0$^\circ$, 10$^\circ$, 20$^\circ$, 30$^\circ$, and 40$^\circ$, while the polynomial coefficients $A_{i}$ and the aperture-meta-optic distance were simultaneously varied to minimize the angle dependent spot sizes (all weighted equally). At the optimal aperture-meta-optic distance of 0.9 cm, the meta-optic diameter must be at least 3 cm to capture all the incident rays through the aperture to achieve a FoV of 80$^\circ$. In other works \cite{Yang23}, a spacer with higher refractive index was used between the aperture and the meta-optic, reducing the spatial separation of beams at larger FoV, thus reducing the necessary diameter of the optic. However, we did not take this approach to maintain a lightweight design. The arrangement of the optimized aperture-meta-optic system is shown in Figure \ref{Fig:Schematic}a. 

To implement the optimized phase profile in an all-silicon (n = 3.47 at 10 $\mu$m \cite{Chan05}) meta-optic, we used square-shaped pillars, which are arranged in a square lattice, schematically shown in Figure \ref{Fig:Schematic}b. Based on parameter sweeps for periodicity, height, and width of the pillar, using rigorous coupled wave analysis (RCWA) \cite{LiuV12}, we set the dimensions to a height of 10 $\mu$m and lattice periodicity of 4 $\mu$m. The phase shift and transmission of the scatterer as a function of width are shown in Figures \ref{Fig:Schematic}c and \ref{Fig:Schematic}d, respectively, for normal and off-axis angles of incidence. We use pillars of widths between 1.7 and 2.4 $\mu$m to achieve the full $2\pi$ phase coverage and excluded the post widths in the grey shaded region to avoid low transmission. We find that the phase dispersion with respect to the incident angle is relatively small as shown in Figure \ref{Fig:Schematic}c. Hence, we used the phase response for normally incident light to map the pillar geometry to the optimized phase distribution.

\section{Fabrication}
\begin{figure}[h!]
\centering\includegraphics[width=12cm]{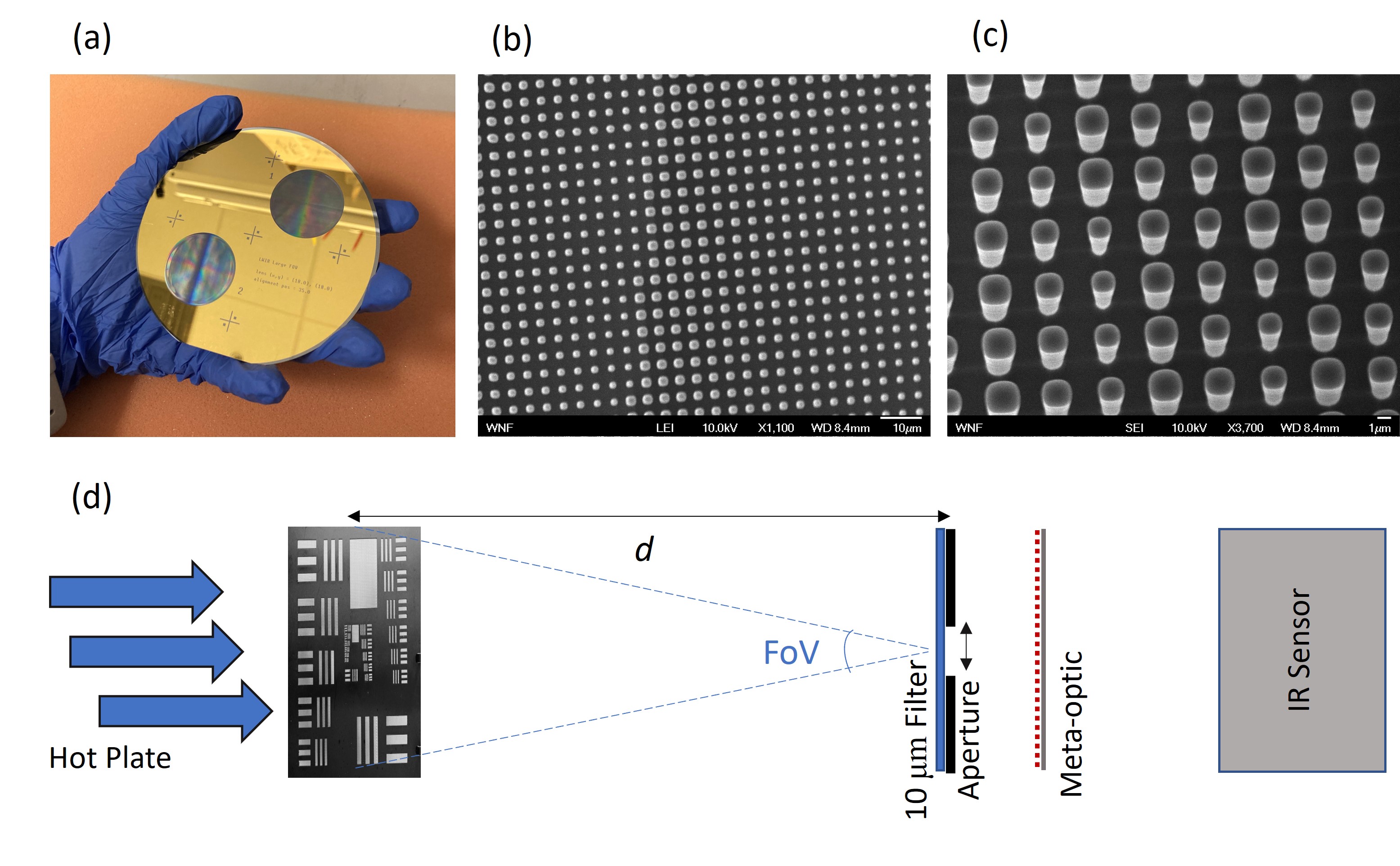}
\caption{Device fabrication and characterization. (a) Photograph of the fabricated device. (b) Top-down scanning electron microscope (SEM) image of the fabricated device, and (c) the same zoomed in and from a slightly oblique viewing angle. (d) Schematic of the experimental setup to characterize the meta-optic.}
\label{Fig:fab_charact}
\end{figure}

We fabricated the 3 cm diameter meta-optic in a 300 $\mu$m thick, double-side polished silicon wafer. First, we deposited an on-chip metal aperture to eliminate stray light passing through the wafer outside the meta-optic. We spin-coated a negative photoresist (NR9G-3000PY) and wrote the aperture pattern by direct laser write lithography (using Heidelberg DWL-66+) and development in AD-10 developer. We then evaporated 10 nm Cr and 100 nm Au via an e-beam evaporator (CHA Solution) and lifted off in acetone to complete the on-wafer aperture. Then, to fabricate the meta-optic pillars, we spin coated a positive photoresist (AZ-1512) and again used direct laser write lithography (Heidelberg DWL-66+) followed by development in AZ-726 developer to pattern the optic. The pattern was then transferred into the underlying silicon using deep reactive ion etching (SPTS Rapier), which was timed to etch $\sim 10$ $\mu $m into the silicon. Finally, oxygen plasma (YES CV200 RFS) was used to strip residual photoresist. A camera image of the fabricated meta-optic wafer is shown in Figure \ref{Fig:fab_charact}a. A scanning electron microscope (SEM) image of the fabricated device from a top-down view is shown in Figure \ref{Fig:fab_charact}b, which shows a section of pillars arranged according to the design. Further inspection at an oblique angle (Figure \ref{Fig:fab_charact}c) shows nearly vertical sidewalls. Both devices on the wafer have an identical design, and were fabricated for redundancy and performed similarly in the experiment. 

\section{Measurements and Results}
The measurement setup is depicted in Figure \ref{Fig:fab_charact}d. A thermal radiation source (hot plate at 65$^\circ$C) was placed in front of a matte black paint-coated aluminum stencil target (the imaging object). For demonstrating image quality, the target pattern was the 1951 USAF resolution chart scaled up by a factor of 2.5, with the cut-out area extending 12.5 cm wide and 11.5 cm tall. Thus, to demonstrate 80$^\circ$ FoV imaging, we placed the object at a distance of 7.45 cm from the aperture of the imaging system such that the light from the outermost feature was 40$^\circ$ relative to normal incidence. The imaging system included the 1 cm external aperture (a thin aluminum sheet with a machined hole), the fabricated meta-optic (Figure \ref{Fig:fab_charact}c), and a FLIR Boson thermal camera core with 12 $\mu$m/pixel resolution. For narrowband characterization, an optical filter (Thorlabs FB10000-500, IR bandpass filter with central wavelength 10.0 $\mu$m and FWHM 0.5 $\mu$m) was placed directly in front of the external aperture. For broadband characterization no filter was placed in front of the aperture and detection was limited to the response of the camera sensor (8 - 14 $\mu$m wavelength). In all imaging experiments, we adjusted the position of the imaging sensor to optimize the image quality. 

\begin{figure}[h!]
\centering\includegraphics[width=12cm]{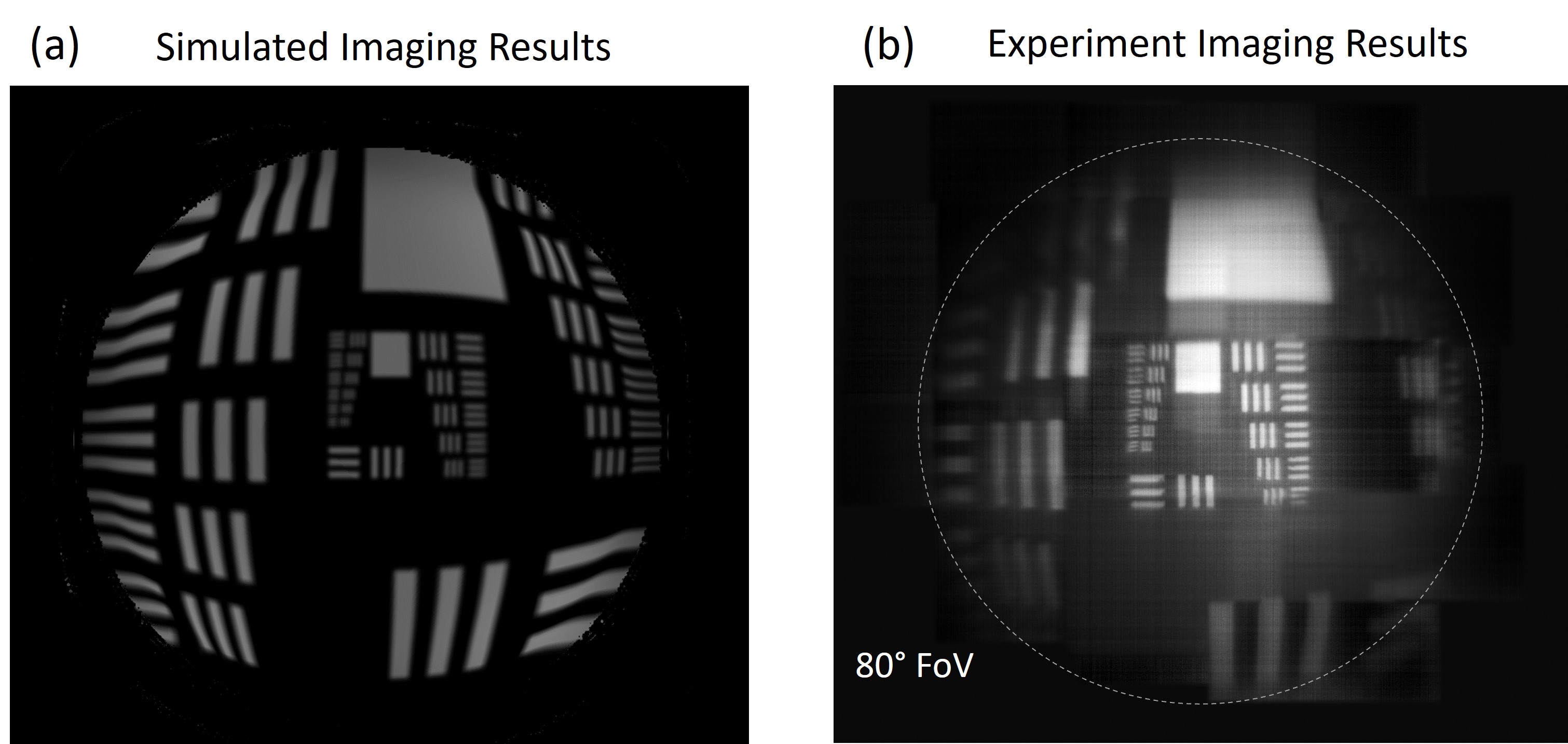}
\caption{Large FoV imaging results. (a) Simulated results via ray tracing through the optimized lens system. (b) Experimental results demonstrating 80$^\circ$ FoV under narrowband illumination.}
\label{Fig:wideFov1}
\end{figure}

Imaging results under narrowband illumination and 7.45 cm object distance, demonstrating the 80$^\circ$ FoV, and a comparison to simulations are summarized in Figure \ref{Fig:wideFov1}. In Figure \ref{Fig:wideFov1}a, ray tracing results (10 $\mu$m) for the optimized lens system are shown, based on an arrangement corresponding to the experimental setup. The experimental capture is shown in Figure \ref{Fig:wideFov1}b for comparison. Because the size of the large FoV image exceeded the size of the camera sensor, we translated the camera over the size of the image, collecting a 5x5 array of images which were stitched together into the final image shown in Figure \ref{Fig:wideFov1}b. In general, a qualitatively good agreement in resolution with simulation is achieved, and the image maintains acceptable quality throughout the entire FoV. In the experimental results, line features on the order of 350 $\mu$m can be resolved at the specific object distance, corresponding to angular resolution of 4.6 mrad. We note that, while the resolution is not diffraction-limited, this is qualitatively good enough for many LWIR imaging applications. The curvature of the image towards 40$^\circ$ is due to barrel distortion or the "fish-eye" effect, which is a correctable distortion that is due to mismatch between the displacement of the focal spot and the paraxial displacement at large FoV \cite{Mart20,Arba16-2,Laik80}. 

\begin{figure}[h!]
\centering\includegraphics[width=12cm]{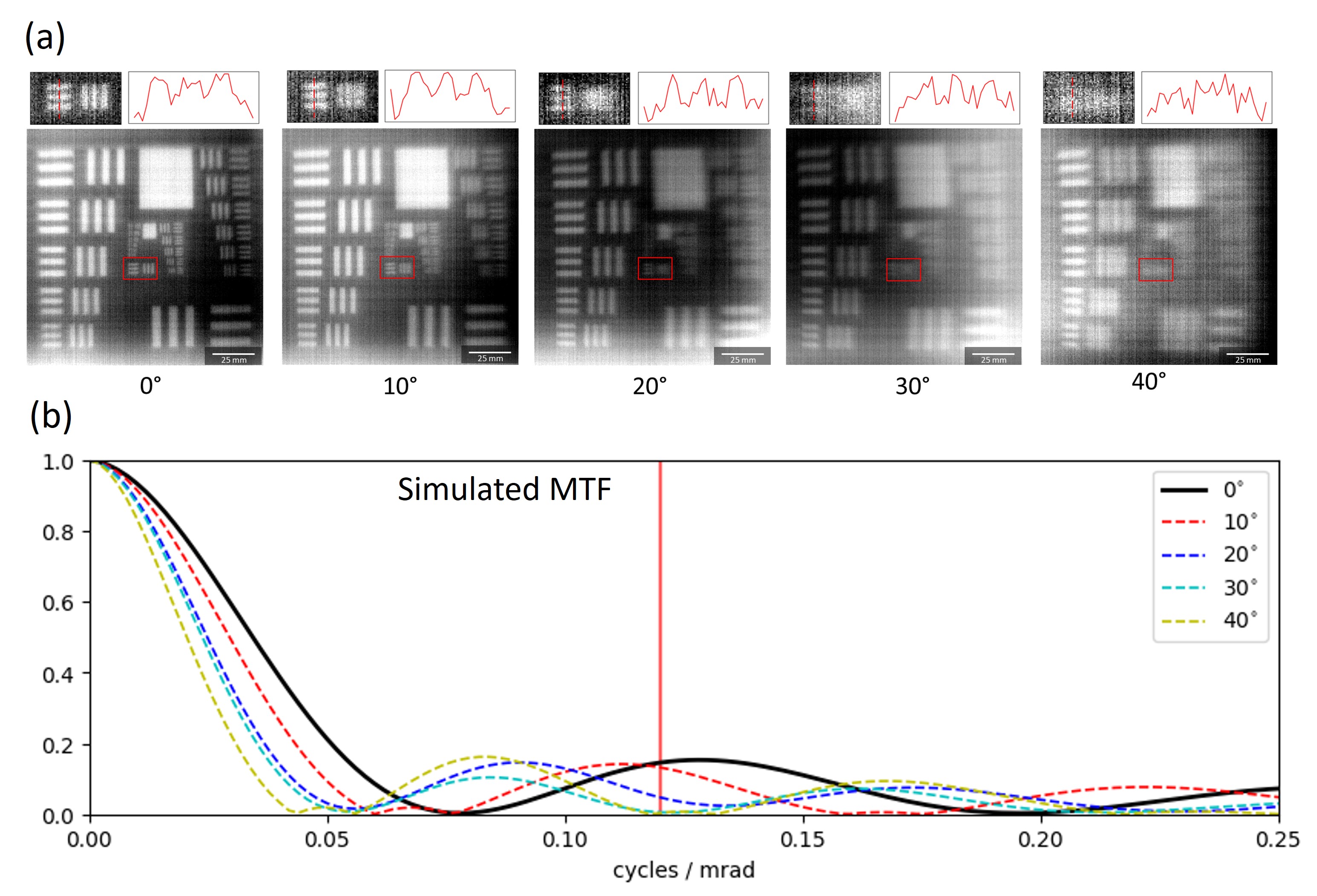}
\caption{Imaging quality at angles of incidence up to 40$^\circ$ half-FoV. (a) Experimental results demonstrating the resolution of an image at incident angles up to 40$^\circ$. The insets show detail of the region denoted by the red square, with a vertical profile taken along the red dashed line. (b) Simulated MTF at incident angles up to 40$^\circ$. The vertical red line indicates the resolution of the groups highlighted by the insets from (a).}
\label{Fig:wideFoV2}
\end{figure}

To further assess the meta-optic performance, we compared the same features across the entire field of view. We note that the thermal light sources are too weak and the camera is too noisy to effectively measure the point spread function (PSF) or modulation transfer function (MTF) of the meta-optic. An LWIR laser source and cooled camera will be needed to measure them. Due to the lack of these in our experiment, we rely on directly imaging the USAF resolution chart, and estimate the resolution. We placed the target at a distance of 30 cm from the imaging system and rotated the entire imaging system relative to the optical axis in increments of 10$^\circ$ up to 40$^\circ$. The results are shown in Figure \ref{Fig:wideFoV2}a. The insets show lines with widths of 1.25 mm, which corresponds to an angular resolution of 4.2 mrad per line, or equivalently, 0.12 cycles / mrad resolution. To quantitatively corroborate the experimental results, we plot the simulated MTF of the imaging system in Figure \ref{Fig:wideFoV2}b, at angles of incidence up to 40$^\circ$. The angular resolution depicted in the insets (0.12 cycles / mrad) is shown by the vertical line. In experiment, we can easily resolve these groups for angles up to 20$^\circ$, while for larger angles, especially at 40$^\circ$, these groups are beyond what can be resolved clearly. We note that some variation in intensity across the images is apparent, which we attribute to expected changes in transmission over various angles of incidence due to the meta-atom scatterer response. In general, as shown in Figure \ref{Fig:Schematic}b, the average transmission decreases for increased angle of incidence, and the transmission at 20$^\circ$ is particularly low, which is qualitatively consistent with the observed results. Consistent with the experimental results, the simulated MTF is slightly higher at normal and small angles of incidence (0$^\circ$ and 10$^\circ$) than it is for larger angles of incidence (20$^\circ$ and beyond). 

Interestingly, the experimental results slightly outperform the simulated results in terms of resolution, both here and in Figure \ref{Fig:wideFov1} upon close inspection of the inner groups. We attribute this to the fact that the simulation assumed collimated input light, whereas the experimental illumination was not collimated, and that we manually adjusted the distances between the aperture, meta-optic, and camera (to within experimental uncertainty) to qualitatively optimize the image quality. To optimize for the human eye, we likely produced slightly better imaging at normal incidence at the expense of slightly worse imaging at larger angles of incidence as compared to the simulation, which aimed to simultaneously optimize for all angles of incidence. The estimated 4.2 mrad experimental resolution was determined from a cut near the center of the image, where we were best able to optimize our experimental conditions. 

\begin{figure}[h!]
\centering\includegraphics[width=12cm]{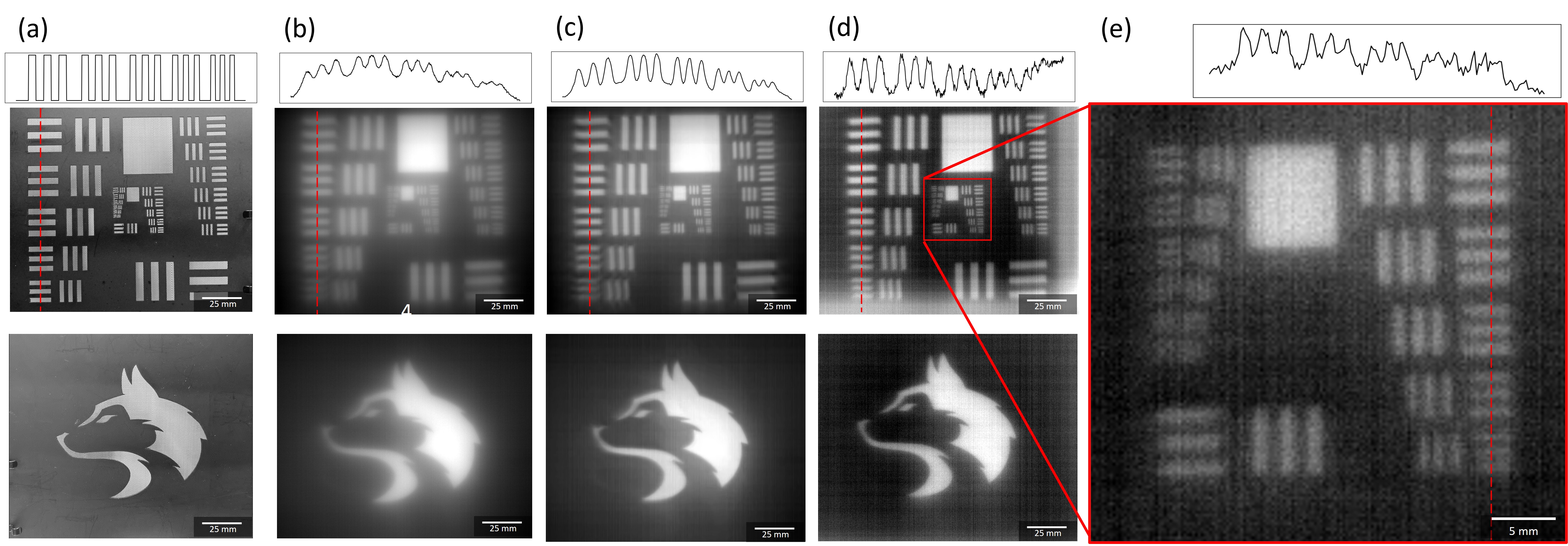}
\caption{Broadband imaging and computational post-processing. (a) Ground truth. (b) Meta-optic image under unfiltered, broadband ambient thermal illumination. (c) The images of (b) with an additional computational postprocessing step applied. (d) Capture with narrowband filter at 10 $\mu$m illumination, without any additional computational processing. The inset (e) shows the resolution of the inner groups of the imaging target.}
\label{Fig:Broadband}
\end{figure}

Finally, we characterized the imaging performance under conditions which exceed the nominal design constraints, namely broadband illumination (8 - 14 $\mu$m) rather than single-wavelength illumination for which the system was particularly designed. Although the system was optimized for a wavelength of 10 $\mu$m, operation over a broader spectral range may be indispensable or simply more practical in realistic applications. However, chromatic aberrations which reduce the image quality under broadband illumination are a challenge in meta-optics systems \cite{Arba17,Shre18,Huan20,Chen18}. Specifically, the focal length of a standard metalens is proportional to 1/$\lambda$, which leads to focal spots at different points along the optical axis for different wavelengths \cite{Arba16}; this effect manifests as a haze in broadband images. To counteract this deterioration in the image quality, computational post processing steps have been shown to be useful \cite{Huan20,Colb18}. We demonstrate this capability for a moderate 30$^\circ$ FoV, which is the largest FoV attainable in a single image due to the limited size of our camera sensor. The results are summarized in Figure \ref{Fig:Broadband} for two test targets, a 2.5 times scaled USAF target and an artistic sketch of a husky; mobile camera images of the objects are shown in Figure \ref{Fig:Broadband}a. First, we show the meta-optic images captured under broadband thermal illumination, without any computational postprocessing, in Figure \ref{Fig:Broadband}b. Notably, a strong haze is apparent in the images, which reduces the image contrast for high resolution features. We then applied a computational post-processing step on the raw images with a Gaussian sharpening filter and bm3d denoising algorithm \cite{Maki20}, which reduced the haze and increased the contrast of the features, as apparent in Figure \ref{Fig:Broadband}c. The image quality was significantly improved after these steps, and we achieve on-par imaging quality with the narrowband imaging at the design wavelength (Figure \ref{Fig:Broadband}d). The inset Figure \ref{Fig:Broadband}e shows the inner groups of the narrowband imaging result to demonstrate detail. The line plots above the USAF images show line cuts as denoted by the dashed red lines for additional clarity.

\section{Discussion}
We demonstrated an all-silicon thermal meta-optic imaging system for 10 $\mu$m wavelength with 80$^\circ$ FoV. While the imaging resolution is not diffraction-limited, acceptable quality is maintained over a large FoV and for relatively large entrance aperture of 1 cm $(\sim1000\lambda)$. In general, the achievable resolution of a single lens is either diffraction-limited, as in the case of a lens with large f-number and small diameter, or limited by geometric aberrations, in the case of a lens with large diameter \cite{Lohm89,Ozak94}. As the diameter increases, geometric aberrations become dominant because distance between rays and the optical axis increases. For an appropriately scaled system (keeping the f-number constant) based on the design in this paper, geometric aberrations become severe beyond 2 cm entrance aperture for normally incident light. However, wide FoV performance places additional constraints on lens designs, and as shown in other works \cite{Mart20} the constraint for diffraction-limited resolutionmust be relaxed to achieve a wide FoV. In an external aperture design like the one discussed in this paper, reducing the entrance aperture improves geometric aberrations for performance closer to the diffraction limit, although the diffraction-limited spot size has also increased due to the increase in f-number. Additionally, larger entrance aperture causes greater spatial overlap between light rays of different angles of incidence, and therefore the lens phase profile cannot be fully optimized for all angles simultaneously. As a result, the imaging resolution of a large FoV, large aperture lens is necessarily reduced from the diffraction limit for all incident angles. In the case of this lens, the geometric aberrations are comparable in severity over the entire FoV. Alternatively, foveated imaging, as recently demonstrated in LWIR \cite{Sara23}, combines a narrow FoV but high-resolution lens and a large FoV, low-resolution lens to achieve high resolution at the center of an image, with low resolution over large FoV for context.

The imaging resolution could be improved by either reducing the entrance aperture or increasing the distance between the aperture and the optic to further spatially separate the incident angles. As the lens is currently designed, there are areas of the lens where the 10$^\circ$ and 40$^\circ$ incident light overlaps, and the lens cannot simultaneously be optimized for both. However, either decreasing the entrance aperture or increasing the aperture-lens distance would change the effective f-number of the system, resulting in fundamentally different characteristics. In addition, reducing the entrance aperture has the negative impact of lowering the light throughput, and subsequently the signal-to-noise ratio. Since many LWIR applications involve imaging distant objects or low-light conditions and limited camera sensitivity, maintaining a large aperture benefits the overall image quality. Increasing the aperture-lens distance would increase the size of the system and require a larger meta-optic, both of which may be impractical for space-constrained systems. However, we still benefit from the light weight of the meta-optics over refractive lenses. Here, we aim to present a well-balanced design, which could be modified to fit specific applications as required by constraints on size, weight, resolution, and FoV. Additionally, we note that extension to even larger FoV is possible by increasing the size of the meta-optic relative to the external aperture to capture rays of larger incident angles.

This meta-optic device is fabricated from a single silicon wafer, which is cost-effective and easily compatible with existing fabrication techniques. Additionally, the high refractive index of silicon (n = 3.47 at 10 $\mu$m) provides high contrast with the surrounding air (n = 1), allowing for compact devices. However, the high contrast also yields relatively high reflections at the interfaces, leading to reduced transmission. Potential ways to improve the performance of this device are to use an anti-reflective coating to reduce these reflections for an increase in overall transmission \cite{Vita22,Nalb22}.

 The next challenge for this design would be to achieve broadband operation over the entire LWIR wavelength range (8 - 12 $\mu$m). While we demonstrate that some degree of broadband imaging is possible despite designing the lens for a single wavelength, a wide FoV lens with inherently broadband operation has not yet been shown in thermal meta-optics. However, recent works have demonstrated high-quality full-color imaging in the visible, which can be adapted for the LWIR regime as well \cite{Huan22}.

In conclusion, we have designed and demonstrated a 1 cm aperture, 80$^\circ$ FoV lens system for LWIR imaging, specifically designed for 10 $\mu$m illumination. By direct measurement, we quantify the angular resolution to be better than 5 mrad, and demonstrate that this resolution degrades only slightly up to incident angles of 40$^\circ$. The light weight and compact nature of these meta-optics together with large FoV functionality can contribute significantly towards the development of highly integrated and lightweight LWIR imaging systems, for applications including night vision goggles, aerial surveillance, and thermal tomography. 
\medskip

\section*{Funding}
\noindent Funding for this work was supported by the federal SBIR program. Part of this work was conducted at a National Nanotechnology Coordinated Infrastructure (NNCI) site at the University of Washington with partial support from the National Science Foundation.


\section*{Disclosures}
\noindent K.B. and A.M. are co-founders of Tunoptix, which is commercializing similar meta-optics in the visible.

\section*{Data availability} 
\noindent Data underlying the results presented in this paper are not publicly available at this time but may be obtained from the authors upon reasonable request.

\bigskip



\bibliographystyle{abbrvnat}
\bibliography{LWIR_LargeFoV}

\end{document}